\documentclass[
aps,prl,
reprint,
a4paper,
superscriptaddress,
floatfix,
]{revtex4-1}
\usepackage[utf8]{inputenc}
\usepackage[T1]{fontenc}

\usepackage{amsmath,amsthm,amsfonts}
\usepackage{braket}
\usepackage{graphicx}
\usepackage{rotating}
\usepackage{hyperref}
\usepackage{units}
\usepackage{xfrac}

\hypersetup{citecolor=magenta}
\hypersetup{colorlinks=true}
\hypersetup{linkcolor=blue}
\hypersetup{urlcolor=blue}

\renewcommand{\doi}[2]{{\href{http://doi.org/#1}{#2}}}
\newcommand{\error}[3]{\left( #1 \pm #2\,\right) \text{#3}}
\newcommand{\val}[2]{#1\, \text{#2}}

\begin{document}


\title{Magic trapping of a Rydberg ion with a diminished static polarizability}

\author{Fabian Pokorny}
\email{fabian.pokorny@fysik.su.se}
\affiliation{Department of Physics,
Stockholm University,
10691 Stockholm, Sweden}

\author{Chi Zhang}
\email{chi.zhang@fysik.su.se}
\affiliation{Department of Physics,
Stockholm University,
10691 Stockholm, Sweden}

\author{Gerard Higgins}
\email{gerard.higgins@fysik.su.se}
\affiliation{Department of Physics,
Stockholm University,
10691 Stockholm, Sweden}

\author{Markus Hennrich}
\email{markus.hennrich@fysik.su.se}
\affiliation{Department of Physics,
Stockholm University,
10691 Stockholm, Sweden}

\begin{abstract}
Highly excited Rydberg states are usually extremely polarizable and exceedingly sensitive to electric fields.
Because of this Rydberg ions confined in electric fields have state-dependent trapping potentials.
We engineer a Rydberg state that is insensitive to electric fields by coupling two Rydberg states with static polarizabilities of opposite sign, in this way we achieve state-independent \textit{magic} trapping.
We show that the magically-trapped ion can be coherently excited to the Rydberg state without the need for control of the ion's motion.
\end{abstract}

\maketitle

{\em Introduction.---}%
Atoms and ions in highly-excited Rydberg states are extremely sensitive to external fields, as well as to nearby Rydberg atoms.
These strong interactions allow for the creation of atom-photon interfaces \cite{Hafezi2012, Pritchard2014, Sarkany2015}, implementation of nonlinear optics at the few-photon level \cite{Firstenberg2016}, engineering of atom-atom interactions for quantum computation and simulation \cite{Saffman2010, Saffman2016, Muller2008}, and for studies of exotic molecular states \cite{Shaffer2018, Eiles2019}.
Their sensitivity to external electric fields enables the creation of long-lived Rydberg Stark states \cite{Cantat-Moltrecht2020}, excellent electric field sensors \cite{Fan2015, Meyer2020}, and near-unity detection of Rydberg states using field ionization.
However, fluctuating stray electric fields cause unwanted fluctuating resonance shifts, and in many experiments stray fields must be carefully monitored and corrected.
This is particularly difficult when the Rydberg atoms are close to surfaces, as is required for the miniturization of experimental setups \cite{Tauschinsky2010, Bernon2013} and for enhanced microwave (MW) coupling using planar waveguides \cite{Hogan2012, Thiele2014, Gard2017}.

In this work we reduce the static polarizability of a Rydberg atomic ion, and show that it becomes insensitive to low-frequency electric fields.
We accomplish this by coupling two Rydberg states with polarizabilities of opposite sign using a MW field \cite{Li2014, Booth2018}.
This allows us to achieve the same trapping potential for an ion in a low-lying electronic (LLE) state and an ion in a Rydberg state, this is called \textit{magic trapping} \cite{Derevianko2010}.

When two different electronic states of an atom or molecule experience different trapping potentials there arises coupling between electronic and motional degrees of freedom, as well as motional heating \cite{Saffman2016}.
These effects are usually unwanted and motivate attempts to minimise trapping potential differences, i.e. to achieve magic trapping \cite{Zhang2011, Boetes2018}.
In systems of neutral atoms and molecules in optical traps, differences between trapping potentials arise from differences in dynamic polarizabilities \cite{Grimm2000}; while in our system of charged Rydberg ions in a Paul trap, differences between trapping potentials arise from differences in static polarizabilities \cite{Muller2008, Higgins2019}.
We achieve magic trapping by diminishing a Rydberg state's static polarizability so that it matches the negligible polarizability of a LLE state.
We show that with magic trapping coupling between electronic and motional degrees of freedom is mitigated and Rabi oscillations between a LLE state and a Rydberg state become insensitive to the ion's motional state.

It is worth noting that coupling of Rydberg states using a MW field was previously used to reduce the polarizability difference between two Rydberg states \cite{Jones2013, Ni2015}.
While this diminishes decoherence on a specific Rydberg transition, reduction of the Rydberg state polarizability to zero stands to be generally applicable in different areas, especially when coherence between Rydberg and LLE states is required.

{\em Theoretical background.---}%
The coupling between two Rydberg states $\left\lbrace\ket{R_S}, \ket{R_P}\right\rbrace$ by a MW field is described by the Hamiltonian (with the rotating wave approximation)
\begin{equation}
H = \frac{\hbar}{2}
\begin{pmatrix}
0& \Omega_{\mathrm{MW}}\\
\Omega_{\mathrm{MW}} & -2\Delta_{\mathrm{MW}}
\end{pmatrix},
\label{eq:coupling_Hamiltonian}
\end{equation}
with coupling strength $\Omega_{\mathrm{MW}}$ and detuning $\Delta_{\mathrm{MW}}$.
The eigenstates of the coupled system are
\begin{equation}
\begin{split}
\ket{+} &= \sin\theta\ket{R_S} + \cos\theta\ket{R_P}, \\
\ket{-} &= \cos\theta\ket{R_S} - \sin\theta\ket{R_P},
\end{split}
\label{eq:dressed_state}
\end{equation}
where the mixing angle $\theta$ is given by
\begin{equation}
\tan\left(2\theta\right) = -\frac{\Omega_{\mathrm{MW}}}{ \Delta_{\mathrm{MW}}},
\label{eq:mixing_angle}
\end{equation}
and the eigenenergies are 
\begin{equation}
E_{\pm} = - \frac{\hbar}{2}\left(\Delta_{\mathrm{MW}} \pm \sqrt{\Omega_{\mathrm{MW}}^2 + \Delta_{\mathrm{MW}}^2}\right).
\label{eq:energy_splitting}
\end{equation}
The polarizabilities of the MW dressed states are the weighted averages of the constituent polarizabilities
\begin{equation}
\begin{split}
\alpha_{+} = \alpha_{R_S} \sin^2\theta  + \alpha_{R_P} \cos^2\theta \, ,\\
\alpha_{-} = \alpha_{R_S} \cos^2\theta + \alpha_{R_P} \sin^2\theta \, .
\end{split}
\end{equation}
When $\alpha_{R_S}$ and $\alpha_{R_P}$ have opposite signs there exist dressed states  $\ket{V_+}$ and $\ket{V_-}$ with vanishing polarizabilities which satisfy
\begin{equation}
\begin{split}
\alpha_{V_+}=0,\quad \tan^2\theta =- \frac{\alpha_{R_P}}{\alpha_{R_S}} ,\\
\alpha_{V_-}=0, \quad \tan^2\theta =- \frac{\alpha_{R_S}}{\alpha_{R_P}} .
\end{split}
\end{equation}

{\em Experimental setup.---}%
We work with a single $^{88}$Sr$^+$ ion, and couple Rydberg states $\ket{R_S}(46S_{1/2}, m_J = -1/2)$ and $\ket{R_P}(46P_{1/2}, m_J = 1/2)$ using a MW field near $\val{122}{GHz}$.
The ratio of their polarizabilities is $\alpha_{R_P}/\alpha_{R_S} = -4.30$ \cite{Weibin_calc}, and we prepare the vanishing-polarizability state $\ket{V_-} = 0.90 \ket{R_S} - 0.43 \ket{R_P}$ using $\Omega_{\mathrm{MW}}/\Delta_{\mathrm{MW}} = 1.26$.

\begin{figure}[t]
\includegraphics[width=\linewidth]{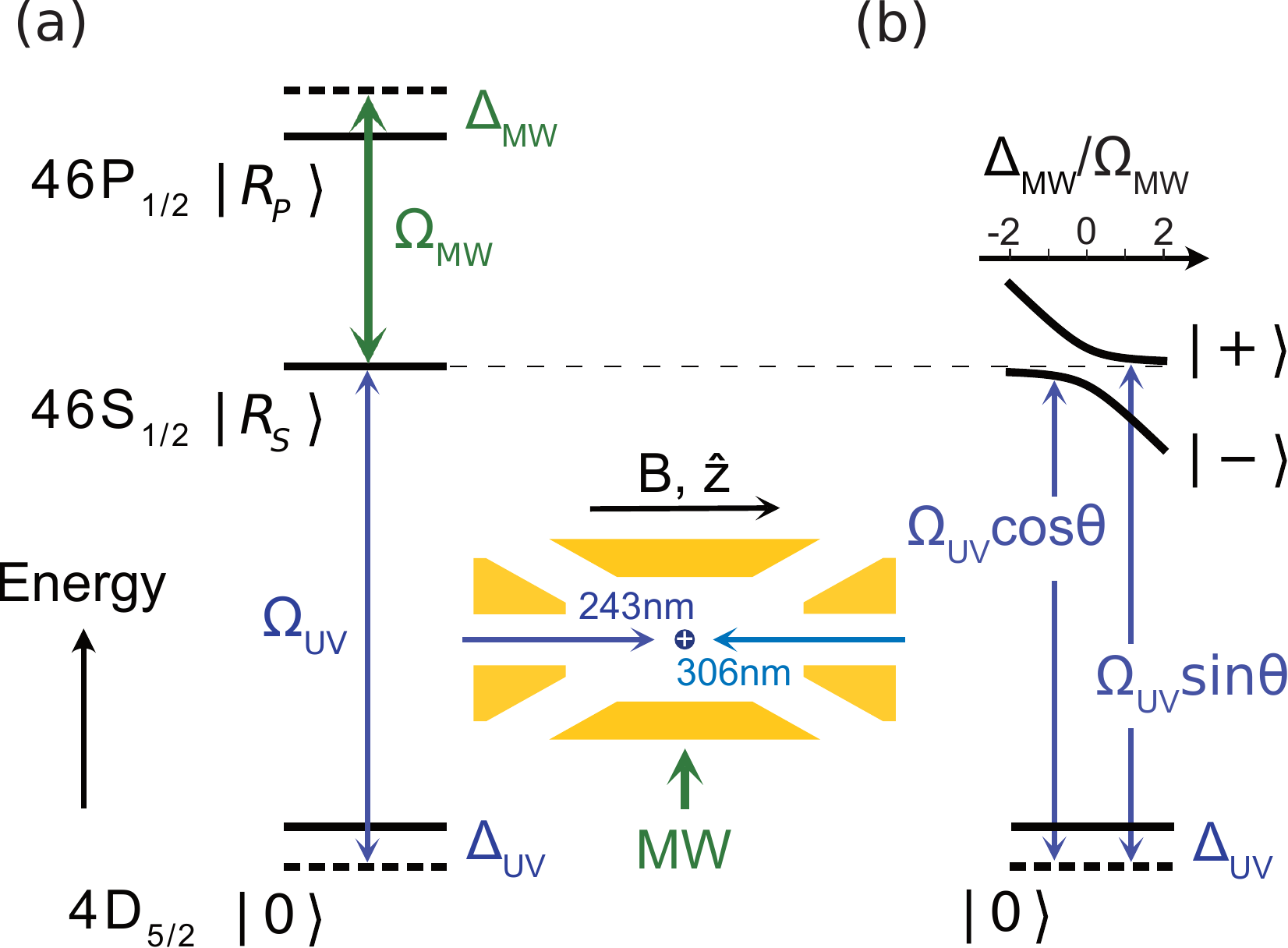}
\caption{\label{fig:experiment} Level scheme of $^{88}$Sr$^+$ showing (a)~the bare states and (b)~the MW dressed states.
Initial state $\ket{0}$ is coupled to Rydberg state $\ket{R_S}$ using a two-UV-photon laser field.
A strong $\val{122}{GHz}$ MW field couples 
 $\ket{R_S} \leftrightarrow \ket{R_P}$, giving rise to the dressed states $\ket{\pm}$.
The inset shows a schematic of the linear Paul trap, the counter-propagating Rydberg-excitation $\val{243}{nm}$ and $\val{306}{nm}$ laser beams, the MW field, the magnetic field direction and the trap symmetry axis $\hat{z}$.}
\end{figure}

The ion is confined in a linear Paul trap, and Rydberg states are excited using a two-UV-photon field which couples LLE state $\ket{0}(4D_{5/2}, m_J = -5/2)$ to $\ket{R_S}$ with coupling strength $\Omega_{\mathrm{UV}}$ and detuning $\Delta_{\mathrm{UV}}$  $\left(\Omega_{\mathrm{UV}} \ll \Omega_{\mathrm{MW}}\right)$. 
The relevant level scheme is shown in Fig~\ref{fig:experiment}. See \cite{Higgins2017a} for details about the experimental setup and the detection of Rydberg ions. For further details about the MW source we refer to the Supplemental Material \cite{Supplemental}.

In ion traps the Rydberg energy levels are modulated by the oscillating trapping fields, both by Stark and quadrupole effects \cite{Muller2008, Higgins2019, Higgins2020}. This strongly affects the $\ket{R_S} \leftrightarrow \ket{R_P}$ coupling when $\hbar \Omega_{\mathrm{MW}}$ is not much larger than the amplitude of the energy level modulation. We work in the regime with $\hbar \Omega_{\mathrm{MW}}$ much stronger than the trap effects.

\begin{figure}[b]
\includegraphics[width=\linewidth]{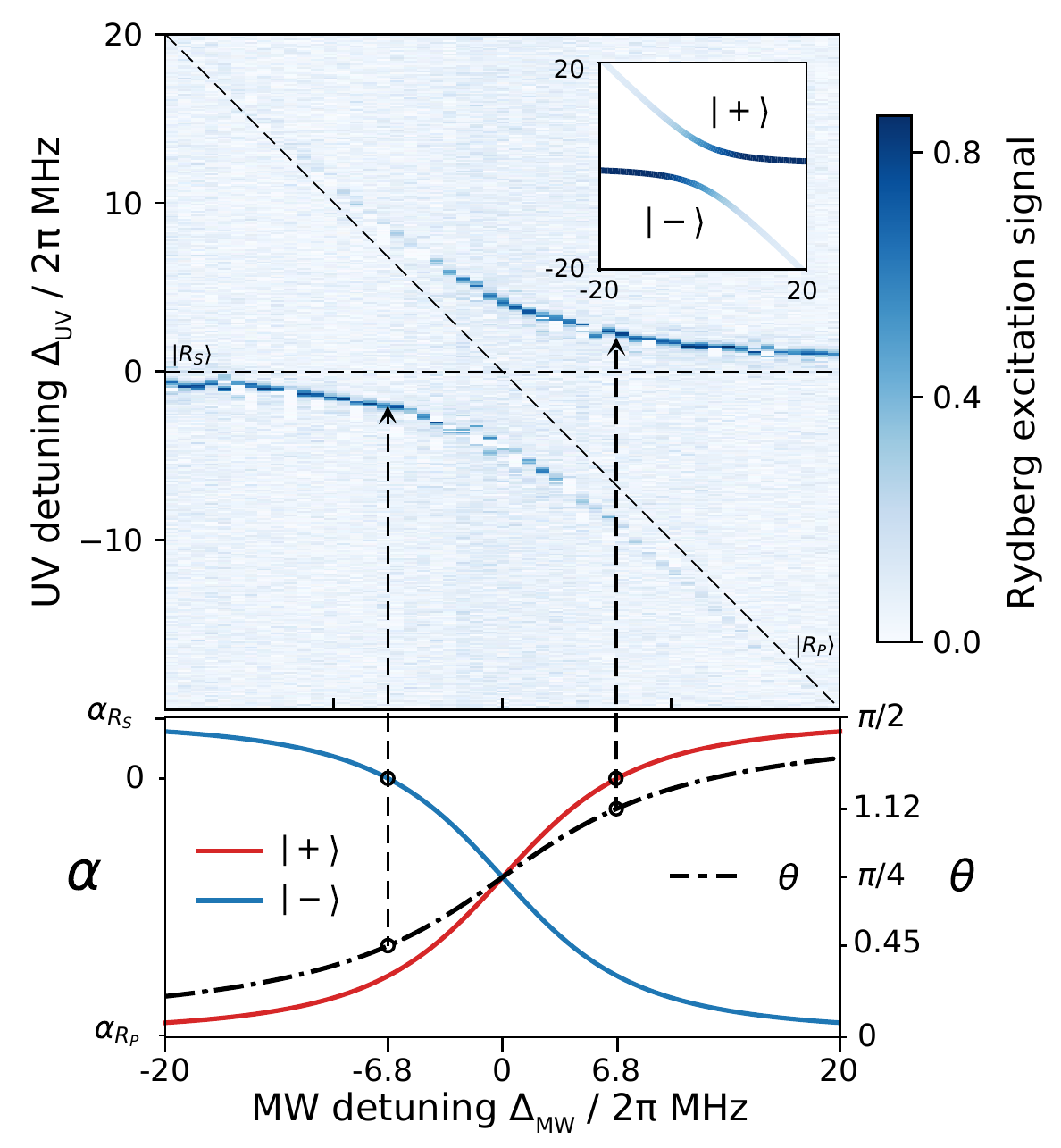}
\caption{\label{fig:ATS}The character of the dressed states $\ket{\pm}$ can be smoothly varied by tuning $\Delta_{\mathrm{MW}}$ (or $\Omega_{\mathrm{MW}}$).
The states are excited by the two-UV-photon field when $\hbar\Delta_{\mathrm{UV}}$ matches the dressed-state energy (Eq.~\ref{eq:energy_splitting}).
The excitation signal is stronger for states with larger $\ket{R_S}$ admixtures since the two-photon field couples $\ket{0} \leftrightarrow\ket{R_S}$.
The bottom part shows the dependence of the dressed state polarizabilities on $\Delta_{\mathrm{MW}}$.
Vanishing polarizability states $\ket{V_{\pm}}$ with $\alpha_{\pm} = 0$ are highlighted.
Here $\Omega_{\mathrm{MW}} = \val{2\pi \times 8.5}{MHz}$.}
\end{figure}

{\em Results.---}%
We probe the MW dressed Rydberg states using spectroscopy as the UV and MW field frequencies are scanned.
The results are shown in Fig.~\ref{fig:ATS}. The observed resonance lines follow Eq.~\ref{eq:energy_splitting}.
The asymptotes at $\Delta_{\mathrm{UV}}=0$ and $\Delta_{\mathrm{UV}}=-\Delta_{\mathrm{MW}}$ correspond to excitation of the bare states $\ket{R_S}$ and $\ket{R_P}$.
The resonance lines display an avoided crossing due to the MW coupling.
For the data in Fig.~\ref{fig:ATS} the mixing angle $\theta$ varies between $\val{0.20}{rad}$ and $\val{1.37}{rad}$ and the vanishing polarizability states $\ket{V_-}$ and $\ket{V_+}$ are excited with $\theta = \val{0.45}{rad}$ and $\theta = \val{1.12}{rad}$, respectively.

\begin{figure}[t]
\includegraphics[width=\linewidth]{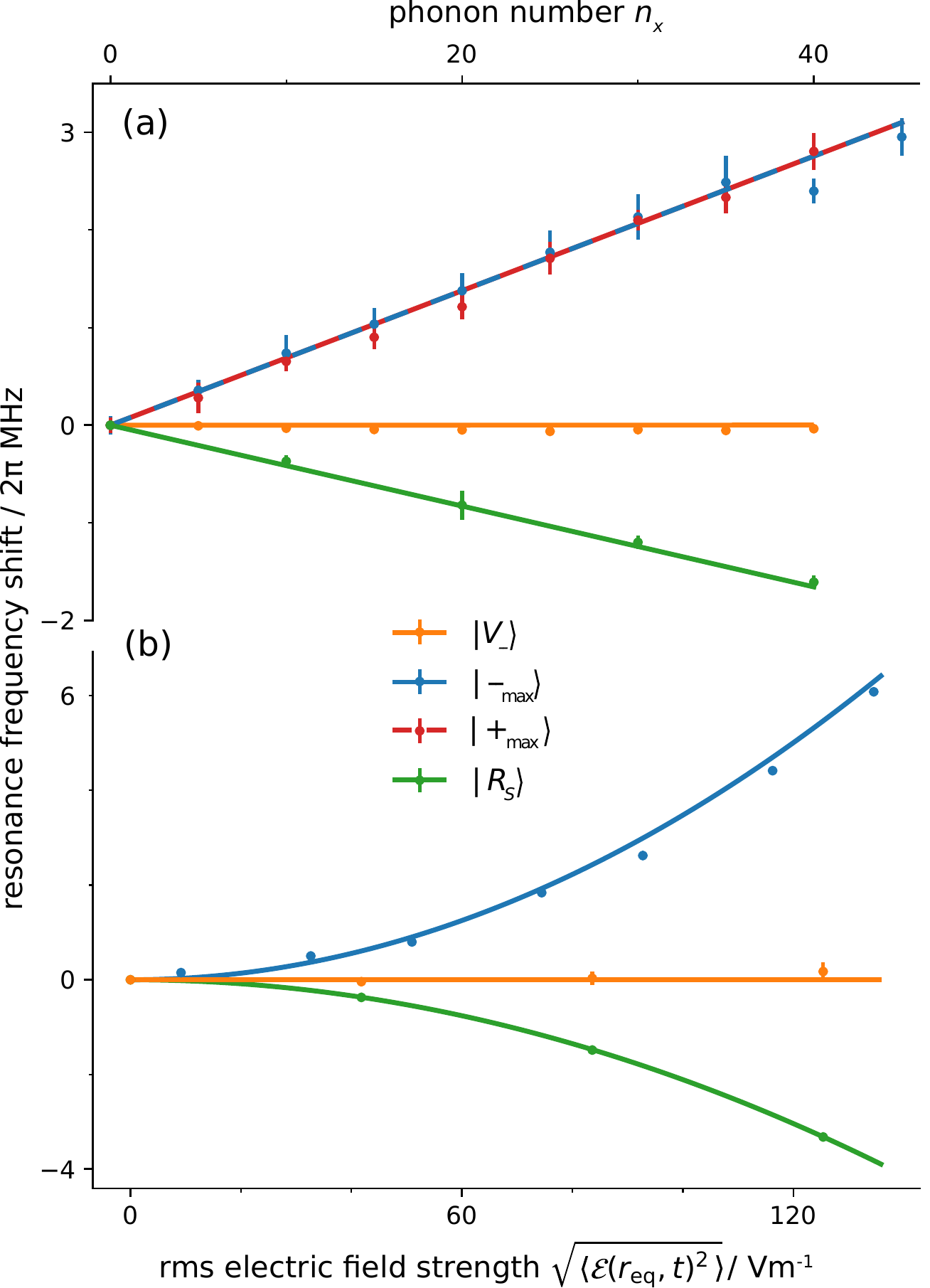}
\caption{\label{fig:Fock}Rydberg states experience strong Stark shifts.
(a)~With increasing phonon number the ion experiences stronger electric fields and Stark shifts increase. 
Maximally-dressed states $\ket{\pm_{\mathrm{max}}}$ show a positive Stark shift while $\ket{R_S}$ shows a negative shift. This confirms that $\alpha_{R_P}$ is larger than and of opposite sign to $\alpha_{R_S}$.
For $\ket{V_-}$, the shift is greatly reduced. 
The trapping frequencies used are $\left\lbrace\omega_x, \omega_y, \omega_z\right\rbrace = 2\pi \times \val{\left\lbrace 1.584, 1.517, 1.020\right\rbrace}{MHz}$.
(b)~Stark shift due to the oscillating electric field at the ion's equilibrium position. $\ket{V_-}$ shows a reduced shift compared to $\ket{R_S}$ and $\ket{-_{\mathrm{max}}}$.
We determine $\mathcal{E}\left(r_{\mathrm{eq}},t\right)$ by probing the micromotion sideband of the narrow transition between ground state and $\ket{0}$ \cite{Berkeland1998}.
Solid lines are calculated shifts.
Error bars represent resonance frequency uncertainties ($68\%$ confidence interval), for some data points they are smaller than the marker size.}
\end{figure}
Next we show that vanishing-polarizability Rydberg states are relatively insensitive to Stark effects.
The electric fields used to confine the ion cause Stark shifts, which depend on the ion's polarizability $\alpha$.
Stark shifts due to trapping fields affect trapped ion precision experiments, including atomic clocks \cite{Rosenband2008, Tamm2014}, as well as experiments which involve highly-sensitive neutral Rydberg atoms \cite{Ewald2019}, Rydberg molecules \cite{Schmid2010}, and Rydberg ions in ion traps \cite{Feldker2015, Higgins2017b, Higgins2019}.
A Paul trap consists of a  combination of static and oscillating electric quadrupole fields.
Ideally the nulls of the static and oscillating fields coincide, and at the ion's equilibrium position the electric field strength is zero.
In this case the ion will still be Stark-shifted due to the non-zero electric field it experiences due the spread of its motional wavefunction which extends beyond the trap center.
This causes the trapping potential to change and the ion's trapping frequencies are altered. For a linear Paul trap the altered frequencies are~\cite{Higgins2019}
\begin{equation}
\omega'_{x,y} \approx \sqrt{\omega_{x,y}^2 - \frac{2\alpha A^2}{M}}, \quad \omega'_{z} \approx \omega_{z}, 
\end{equation}
where $\omega_{x,y,z}$ are the trapping frequencies in the absence of the Stark effect, $x$ and $y$ are radial directions, $z$ is the axial direction, $A$ is the strength of the oscillating quadrupole field and $M$ is the ion mass.
This equation describes usual experimental settings in which the oscillating quadrupole field is much stronger than the static quadrupole field.

As a result the energy required to excite an ion from a LLE state with negligible polarizability to a highly polarizable Rydberg state is Stark shifted depending on the number of phonons in the radial motional modes $n_x, n_y$
\begin{equation}
\begin{split}
\Delta E_{\mathrm{LLE}\rightarrow R} &= \left(n_x+\tfrac{1}{2}\right)\hbar\left(\omega'_x - \omega_x \right) \\
&+ \left(n_y +\tfrac{1}{2}\right)\hbar\left(\omega'_y - \omega_y\right).
\end{split}
\label{eq:energy_fock_state}
\end{equation}
The linear relation between the Stark shift and the phonon number can be understood as follows:
The root mean square (rms) ion position grows as $\sqrt{n_x}$ $\left(\sqrt{n_y}\right)$, the rms electric field strength grows linearly with distance from the trap center, and the Stark shift grows with the square of the electric field strength.

We probe this effect by measuring the change of the Rydberg-excitation frequency as $n_x$ is varied ($n_y$ is kept at zero).
The results are shown in Fig.~\ref{fig:Fock}(a).
Rydberg state $\ket{R_S}$ and the maximally-dressed states $\ket{\pm_\mathrm{max}} = \frac{1}{\sqrt{2}}\left(\ket{R_S} \pm \ket{R_P}\right)$ have large polarizabilities and display large Stark shifts ($\sim \val{50}{kHz per phonon}$).
We tune the MW coupling to achieve Rydberg state $\ket{V_-}$ which shows greatly-diminished Stark shifts.
For $\ket{R_S}$ the altered trapping frequency has a fractional change $\left(\omega_x - \omega'_x\right)/\omega_x = \val{2.6}{\%}$, while $\ket{V_-}$ changes by $\left(\omega_x - \omega'_x\right)/\omega_x = \error{0.145}{0.013}{\%}$, and thus the trapping potential difference between LLE states and $\ket{V_-}$ is diminished -- magic trapping is achieved. We estimate the residual polarizability  $\alpha_{V_-} = \error{0.056}{0.005}{$\alpha_{R_S}$}$.

When the nulls of trap's static and the oscillating quadrupole fields do not coincide the ion experiences an oscillating electric field at its equilibrium position $\mathcal{E}\left(r_{\mathrm{eq}},t \right)$, which causes an additional Stark shift \cite{Higgins2019}
\begin{equation}
\delta = -\tfrac{1}{2}\alpha\langle \mathcal{E}\left(r_{\mathrm{eq}},t \right)^2\rangle ,
\label{eq:Stark_shift}
\end{equation}
where $\langle \, \rangle$ denotes the time-average over an oscillation period ($\val{55}{ns}$ in our system).
Usually ion trappers aim to overlap the static and the oscillating quadrupole field nulls in order to minimize excess micromotion \cite{Berkeland1998}.
We measure the dependence of the Rydberg excitation frequency on the rms electric field strength $\sqrt{\langle\mathcal{E}\left(r_{\mathrm{eq}},t\right)^2\rangle}$, the results are shown in Fig.~\ref{fig:Fock}(b).
The resonance frequency shift scales as $\langle\mathcal{E}\left(r_{\mathrm{eq}},t\right)^2\rangle$, according to Eq.~\ref{eq:Stark_shift}.
The curvature depends on the Rydberg state's static polarizability $\alpha$.
The vanishing polarizability state $\ket{V_-}$ displays a greatly-diminished response compared to $\ket{R_S}$ and $\ket{-_{\mathrm{max}}}$.
The relative curvatures indicate $\alpha_{V_-} = \error{0.054}{0.016}{$\alpha_{R_S}$}$.
\begin{figure}[t]
\includegraphics[width=\linewidth]{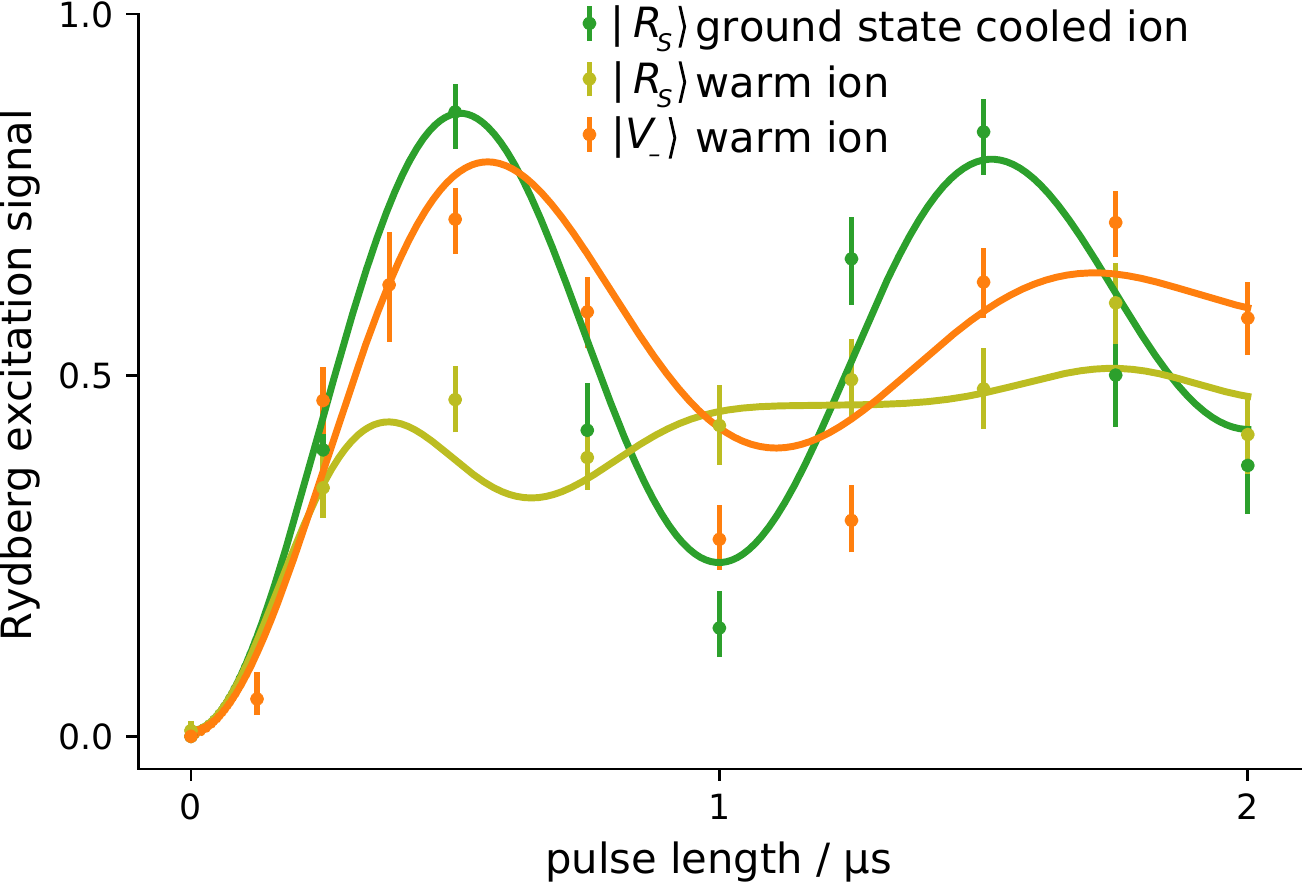}
\caption{\label{fig:Rabi} Rabi oscillations between low-lying electronic state $\ket{0}$ and different Rydberg states.
Rabi oscillations are only observed between $\ket{0}$ and $\ket{R_S}$ for a ground-state cooled ion; for a relatively warm (Doppler cooled) ion the oscillations are washed out because of Stark broadening.
Rabi oscillations between $\ket{0}$ and $\ket{V_-}$ are driven even when the ion is warm, since $\ket{V_-}$ is insensitive to electric fields.
The reduced Rabi-frequency of $\ket{V_-}$ is a consequence of the dressed state's reduced $\ket{R_S}$ component.
Solid lines show simulated results with experimental parameters, using the python module QuTiP \cite{Johansson2013}.
Error bars represent quantum projection noise ($68\%$ confidence interval).}
\end{figure}

The dependence of the Rydberg resonance frequency on the ion's phonon number [Fig.~\ref{fig:Fock}(a)] means the electronic and motional degrees of freedom are correlated, and Rydberg resonances are broadened for an ion with a broad thermal phonon distribution \cite{Higgins2019}.
In previous works we mitigated this Stark broadening by using ground state cooling to narrow the phonon number distribution. This allowed us to coherently excite highly-polarizable Rydberg states \cite{Higgins2017b, Higgins2019} to conduct a Rydberg-entangling gate \cite{Zhang2020}.
In large ion crystals efficient ground-state cooling is impractical and an alternative approach must be followed to avoid the Stark broadening of Rydberg resonances.
Here we demonstrate that Rydberg states can be coherently excited without the need for ground state cooling by reducing the Rydberg state polarizability, to achieve magic trapping.
Figure~\ref{fig:Rabi} shows that Rabi oscillations can be driven between $\ket{0}$ and $\ket{V_-}$ for a relatively warm, Doppler-cooled ion.
Rabi oscillations between $\ket{0}$ and $\ket{R_S}$ however are only visibile when the ion is ground-state cooled; for a warmer, Doppler-cooled ion with a broad phonon-number distribution the oscillations are smeared out.

{\em Conclusion.---}%
We have used MW dressing to diminish the static polarizability of a Rydberg state, thus diminishing the state's sensitivity to low-frequency electric fields.
In this way we reduce the differences between the trapping potentials of a LLE state and a Rydberg state, to achieve magic trapping.
This allows us to coherently excite Rydberg states without careful control of the ion's motional state.
With this method the recent Rydberg ion entangling gate \cite{Zhang2020} can be carried out in long ion strings \cite{Zhang2020b}, in which cooling to the motional ground state is unfeasible.

The reduction of a Rydberg state's polarizability we demonstrate is directly applicable in a range of Rydberg atom experiments in which care is taken to minimise unwanted Stark effects.
For instance, in Rydberg atom quantum computation and simulation experiments, reducing the sensitivity to electric fields mitigates dephasing errors caused by fluctuating Stark shifts.
Of course one must keep in mind that as well as altering the polarizability, MW dressing will affect other Rydberg state properties, such as the lifetime; further it introduces a rotating dipole moment which enables direct dipole-dipole interactions between Rydberg atoms \cite{Muller2008, Zhang2020}.

{\em Acknowledgements.---}
We thank Weibin Li for theory values of Rydberg state energies and polarizabilities.
This work was supported by the Swedish Research Council (Trapped Rydberg Ion Quantum Simulator), by QuantERA project ``ERyQSenS'',
and by the project ``Photonic Quantum Information'' (Knut and Alice Wallenberg 
Foundation, Sweden).


\begin{thebibliography}{99}
\bibitem{Weibin_calc}%
Theory values for Rydberg state energies and polarizabilities were calculated by Weibin Li.
\end{thebibliography}


\appendix
\section{Appendix: Microwave setup}
\begin{figure}
\includegraphics[width=\linewidth]{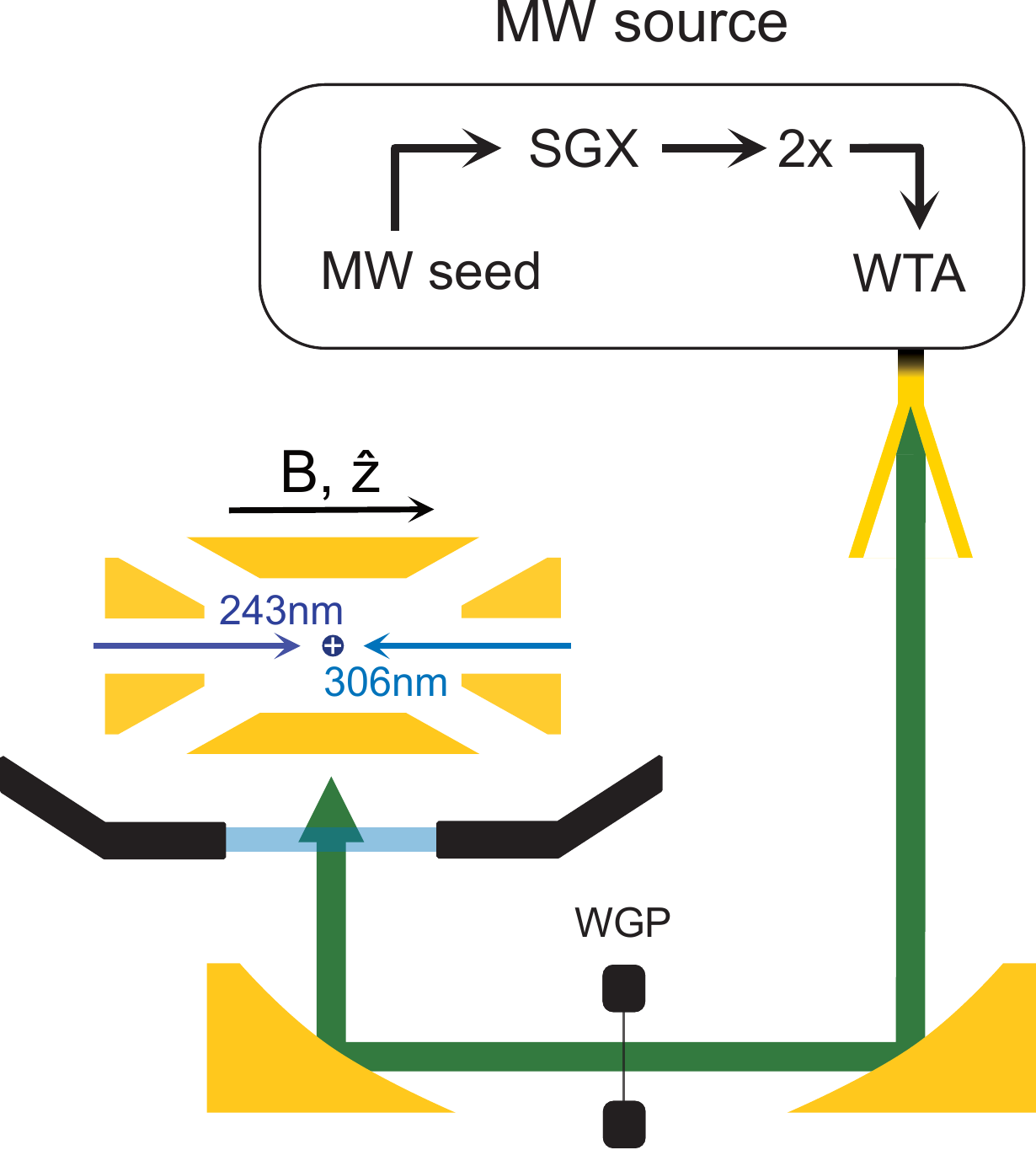}
\caption{\label{fig:MW_setup}Schematics of the MW setup and trap. The MW source consists of a MW seed that gets multiplied (first in a SGX unit and then by a doubler) and attenuated before being coupled into free-space with a conical horn antenna. Two off-axis parabolic mirrors guide the field through a view-port into the trap, a wire-grid polarizer (WPG) removes polarization imperfections.}
\end{figure}

We generate the MW field outside the vacuum chamber through frequency multiplication of a seed signal. The field is coupled to free-space and guided to the ion position through a view-port (see Fig. \ref{fig:MW_setup}).

The seed signal is provided by an Anritzu MG3690C signal generator with an output frequency of $\val{2-20}{GHz}$ and an output power of $\val{10}{dBm}$.
This seed signal feeds a WR15SGX signal generator extension module (SGX) from Virginia Diodes Inc. 
The SGX multiplies the input signal by a factor of 4 and simultaneously amplifies it; the resulting output signal is $\val{50-75}{GHz}$ at $\val{20}{dBm}$. 
An additional WR8.0X2 frequency doubler (Virginia Diodes Inc)
finally results in a field of frequency range of $\val{90-140}{GHz}$ at $\val{10}{dBm}$ output power.

With a WR0.8CH conical horn antenna the MW field is coupled into free space. 
Two off-the-shelf gold-coated off-axis parabolic mirrors (Thorlabs, Edmund Optics) first collimate and then focus the MW field through a fused silica view-port into the ion trap. 
The MW field enters the trap perpendicular to the magnetic field and the UV laser beams.
Through rotation of the MW source around its axis we can change the polarisation of the field between horizontal and vertical, thus changing between $\pi-$ and $\sigma-$ transitions. 
A $\val{15}{$\mu$m}$ wire grid polarizer (PureWave Polarizers) placed in the collimated beam removes polarization imperfections.

From the diameter of the conical horn antenna we use geometric optics to estimate the waist of the $\approx\val{120}{GHz}$ MW field to be $\val{10}{mm}$. 
We measure the transmission of the viewport to be $\approx \val{70}{\%}$. 
Manufacturers do not provide data for the reflectivity of gold coatings in the $\val{100}{GHz}$ frequency range. 
However, literature suggests that metal-coated mirrors readily available from standard suppliers perform adequately in the frequency range we are operating \cite{Parshin2005, Naftaly2011}; we confirm these findings. 
We note that the separation between the trap's blade electrodes is $\val{700}{$\mu$m}$ and the wavelength for $\val{120}{GHz}$ radiation is $\approx\val{2.5}{mm}$.

We find that the intensity of the MW field inside the trap is high enough to drive the $\ket{46S_{1/2}} \leftrightarrow \ket{46P_{1/2}}$ transition with $\Omega_{\mathrm{MW}} >\val{2\pi \times 500}{MHz}$. 
Such high Rabi-frequencies couple even far off-resonant Rydberg states significantly. 
In order to diminish such unwanted coupling we attenuate the MW field with a WTA 90-140 waveguide tunable attenuator from Radiometer Physics. This attenuator is placed between doubler and conical horn to prevent underpumping the doubler.
The maximum attenuation is $\val{40}{dB}$, resulting in a $\Omega_{\mathrm{MW}} \sim \val{2\pi \times 8}{MHz}$.

\end{document}